\documentclass[a4paper,aip,jcp]{revtex4-1}

\usepackage{mathpazo}
\usepackage{amssymb}
\usepackage[fleqn]{amsmath}
\usepackage{graphicx}
\usepackage{subfig}
\usepackage{epstopdf}
\usepackage{fancyhdr}
\usepackage{titlesec}
\usepackage{fncychap}
\usepackage[titletoc,title]{appendix}
\usepackage{framed}
\usepackage[bottom]{footmisc}
\usepackage{bm}

\renewcommand{\eqref}[1]{eq.~\ref{#1}}

\newcommand{\ke}[1]{\vert #1  \rangle}

\begin{document}

\title{Photoinduced molecular chirality probed by ultrafast resonant X-ray spectroscopy}

\author{J\'er\'emy R. Rouxel}
\email{jrouxel@uci.edu}
\author{Markus Kowalewski}
\email{mkowalew@uci.edu}
\author{Shaul Mukamel}
\email{smukamel@uci.edu}
\affiliation{Department of Chemistry, University of California, Irvine,
California 92697-2025, USA}
\date{\today}%

\begin{abstract}
Recently developed circularly polarized X-ray light sources can probe ultrafast chiral electronic and nuclear dynamics through spatially localized resonant core transitions. 
We present simulations of time-resolved circular dichroism (TRCD) signals given by the difference of left and right circularly polarized X-ray probe transmission following an excitation by a circularly polarized optical pump with  variable time  delay. 
Application is made to formamide which is achiral in the ground state and assumes two chiral geometries upon optical excitation to the first valence excited state. Probes resonant with various K-edges (C, N and O) provide different local windows onto the parity breaking geometry change thus revealing enantiomer asymmetry.
\end{abstract}

\maketitle

\section{Introduction}

This article is dedicated to the memory of Ahmed H. Zewail whose inspiring work has pioneered the field of femtochemistry.
Stereochemistry is of crucial importance for biological processes and for chemical
syntheses of natural products. Enantioselective synthesis is a major challenge in organic
chemistry, while discerning and identifying enantiomers is a problem for spectroscopy.
A widely used method for measuring the enantiomer excess is circular dichroism (CD) \cite{berovaCD}; the difference in absorption between left and right polarized light. 
In contrast to conventional linear absorption spectroscopy which is dominated by electric dipole transitions, magnetic dipole transitions are essential in CD spectroscopy.
Time-resolved CD can be used to measure molecular chirality variations on a femtosecond
timescale\cite{CHIR:CHIR20179} and follow the formation and decay of enantiomers on the intrinsic timescale of the molecule. The use of X-ray radiation instead of IR of UV light allows
to measure element specific transitions\cite{bennett2016multidimensional} and thus more specifically address chiral centers
in a molecule.

Bright and coherent X-ray radiation, generated by free electron lasers (XFEL)\cite{PhysRevSTAB.12.060703,PhysRevSTAB.16.110702,
Altarelli20112845} and high harmonic generation (HHG)\cite{nphoton.2010.256} tabletop sources, has paved the way for core resonant ultrafast nonlinear X-ray spectroscopy.
Measuring chirality specific signals requires an optical pump and a X-ray
probe setup \cite{zewail2000femtochemistry} with circularly polarized laser light. Such pulses are now
available at facilities like the Stanford Linear Accelerator Center \cite{higley2016femtosecond} or the Fermi free electron laser at Elettra Sincrotrone Trieste \cite{allaria2014control}.
Circularly polarized X-ray pulses can utilize the element and orbital specificity of X-ray transitions to probe matter chirality thus providing new local windows into molecular geometry changes.

Picosecond circularly polarized X-ray light with relatively
low brightness generated by insertion devices \cite{1.1140915} in synchrotron radiation \cite{SASAKI199487} has been used to study magnetic properties of matter through X-ray magnetic circular dicroism \cite{vanderLaan201495} whereby a CD spectrum is measured in the presence of an external magnetic field which breaks the mirror symmetry.
CD of amino-acids with XUV light has been predicted\cite{Takahashi2015109}.

Two approaches may be employed to measure ultrafast chirality in the X-ray regime. 
The first, chiral HHG (cHHG)\cite{doi:10.1038/nphys3369,nphoton.2014.293}, uses an intense mid-IR field excitation\cite{1508.02890} to ionize a molecule. The released electron is then accelerated in the intense laser field until it recombines with the molecule, emitting HHG light in the process. Enantiomers were found to have a different HHG spectrum depending on the incoming laser ellipticity\cite{doi:10.1038/nphys3369}.
The second technique is CD.
Some dynamics is initiated by optical excitation and the resulting time-dependent chiral signal is then detected
\cite{CHIR:CHIR20179,doi:10.1038/nphys3369,doi:10.1021/cr800268n} by 
the difference in the absorption of left and right polarized resonant X-ray pulses
\cite{10.1063/1.4961470,Ferrari2015}.
Thanks to the strong localization of the core orbitals, this signal should be particularly sensitive  to the local breaking of the mirror symmetry in the vicinity of the selected atom.
The HHG  signal is robust and the first approach is easier to implement with current technology and was investigated both experimentally  and theoretically\cite{doi:10.1038/nphys3369}.
However, the interpretation is not easy  due to the complex multistep nature of the HHG  process.
X-ray CD is harder to measure but easier to interpret.

In this article, we explore computationally this optical pump and X-ray probe CD setup. 
Such time-resolved chirality measurements have so far been limited to the visible and near UV range and to the picosecond timescale\cite{1.4948943,doi:10.1021/jp207270s,LPOR:LPOR201200065}. 
A core resonant X-ray probe can measure faster processes and is more sensitive to the local change of conformation within the molecule due to the element specificity of the X-ray core transition  for atoms located in the vicinity of the chiral center.

\begin{figure}[!h]
  \centering
  \includegraphics[width=0.8\textwidth]{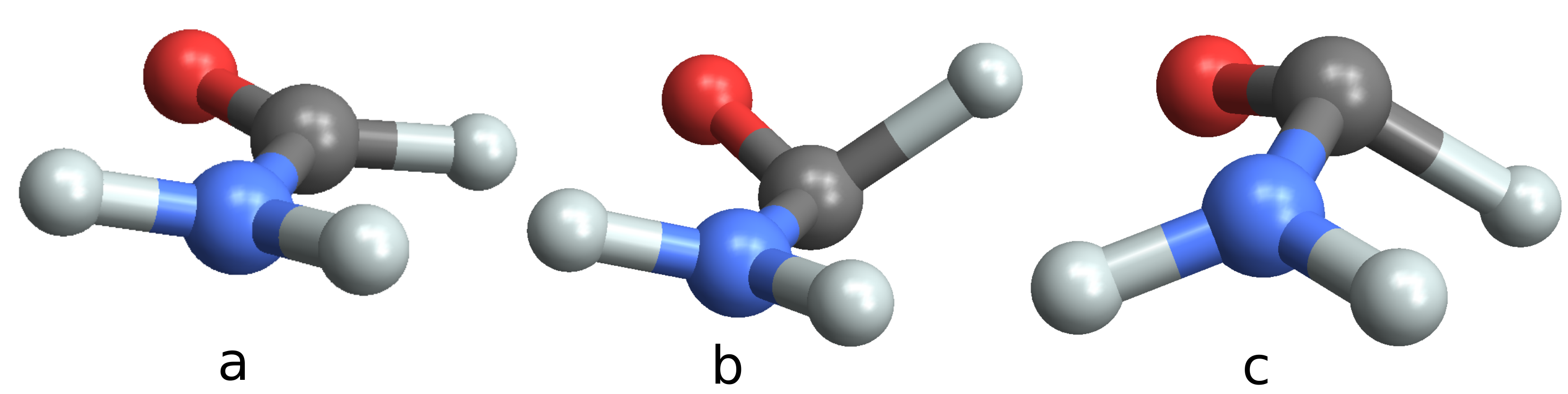}
  \caption{Geometries of formamide in the planar achiral ground state, a, and the two enantiomers in the excited state, b and c.
  \label{fig0}}
\end{figure}

We apply this technique to formamide which is achiral in its ground state. Upon near UV (NUV) excitation an electron from the oxygen lone pair is promoted to the $\pi^*$ bond
of the CO bond. This leads to pyramidalization in the CHO group, creating
a chiral non-planar configuration with two possible enantiomers \cite{FISCHER1979136} as shown in Fig. \ref{fig0}.
Our goal is to probe the $\sim$120 fs geometry change in the excited state and the time evolving chirality through the difference between the absorption of left and right circular X-ray probe polarization. 
Formamide is a good candidate for this study: it contains three soft X-ray chromophores (C, N and O) and the chiral isomerization happens on a femtosecond timescale.

\begin{figure}[!h]
  \centering
  \includegraphics[width=1.\textwidth]{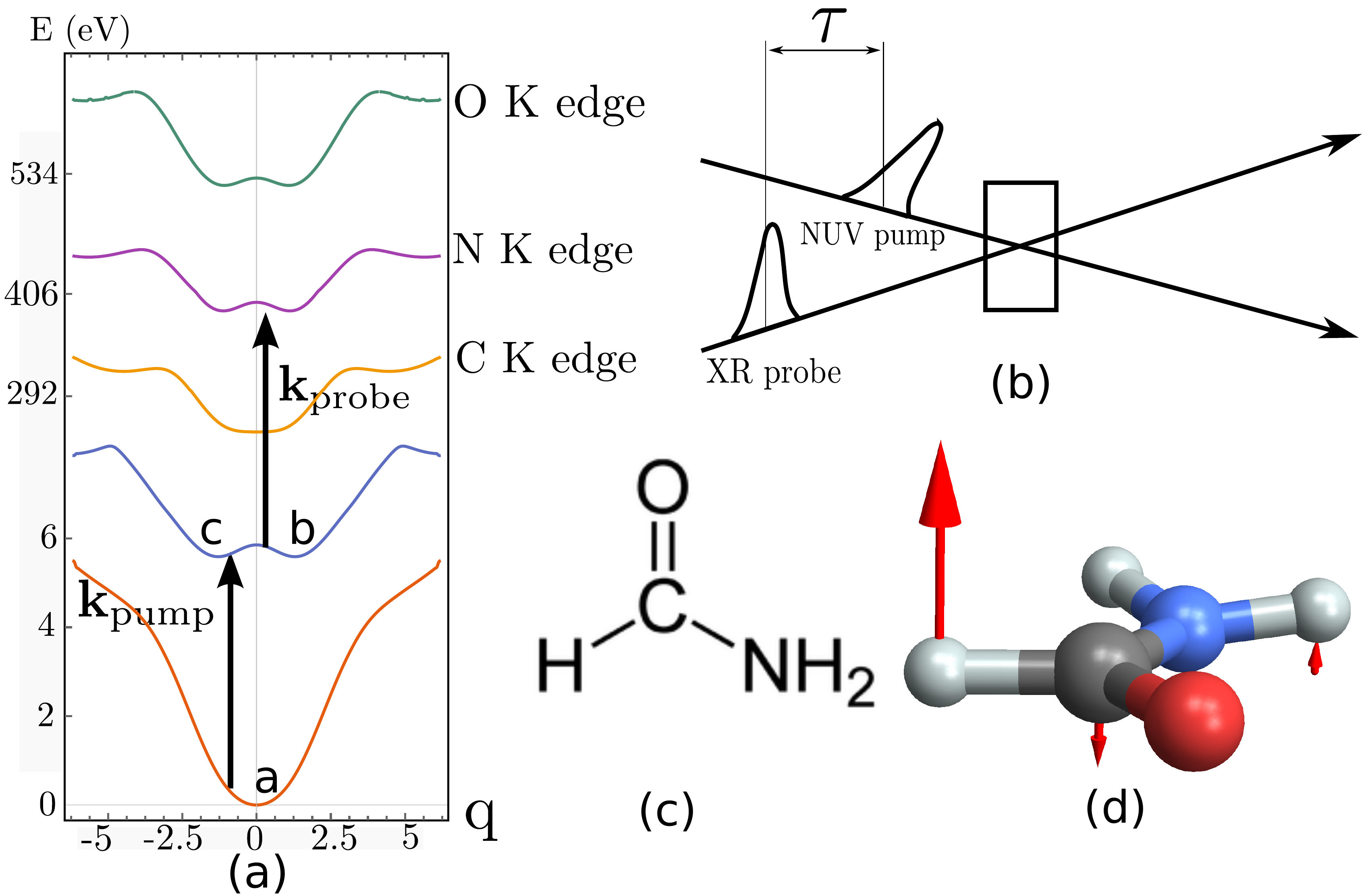}
  \caption{(a) Relevant potential energy surfaces of formamide (chemical structure displayed in (c)) along the out-of-plane bending normal coordinate $q$ initiated by the pump pulse and displayed in (d). 
The calculated potentials for the ground state, the first valence excitation and the C, N and O K-edges are shown. $q=0$ is the planar achiral geometry and the two minima at $\pm 0.6 q$ correspond to the two enantiomers. The geometries a, b and c displayed in Fig. \ref{fig0} are indicated in (a). In the pump-probe scheme sketched in (b), a left polarized NUV pump creates a valence excitation and the molecule then evolves in the excited state double well potential and is then probed after a delay $\tau$ by circularly polarized X-ray light at various K-edges (C, N and O). The difference between left and right probe absorption gives the chiral contribution to the signal.
  \label{fig1}}
\end{figure}

Section \ref{section2} presents the expressions for the transient CD signals. Section \ref{section3} reports the C, N and O K-edge TRCD signals which monitor the bending dynamics upon a valence excitation.
Section \ref{section4} discusses the information obtained from time-resolved X-ray CD.

\section{The time-resolved X-ray circular dichroism signal}
\label{section2}
The TRCD signal is given by the difference in the absorption spectrum of a left and right circularly polarized probe following an excitation with a circularly polarized optical pump pulse. 
The total Hamiltonian of the system is given by 
\begin{equation}
H = H_0 + H_{\text{pu}}(t) + H_{\text{pr}}(t)
\label{hamiltonian}
\end{equation}
with $H_0$ represents the free molecule and 
\begin{eqnarray}
H_{\text{pu}}(t) &=& - \bm\mu\cdot \bold E_{\text{pu}}(t)- \bold m\cdot \bold B_{\text{pu}}(t)\label{hamiltonian2}\\
H_{\text{pr}}(t) &=& - \bm\mu\cdot \bold E_{\text{pr}}(t)- \bold m\cdot \bold B_{\text{pr}}(t)
\end{eqnarray}
\noindent represent the interaction with the pump and the probe. Here, $\bm\mu$ and $\bold m$ are the electric and magnetic dipoles respectively, $\bold E$ and $\bold B$ are the electric and magnetic fields. The electric quadrupole interaction is not included because it gets cancelled out in an isotropic average\cite{barronROA}.
Throughout this article, we consider circularly polarized fields of the form
\begin{eqnarray}
\bold E^{L/R}(t) &=& a(t) \bold e_{L/R}\\
\bold B^{L/R}(t) &=& a(t) \bold b_{L/R}
\end{eqnarray}
\noindent where $\bold e_{L/R}$ and $\bold b_{L/R}$ are the polarization unit vectors of a left or right polarization for the electric and the magnetic fields respectively. We further assume Gaussian field amplitudes
\begin{eqnarray}
a_{\text{pu}}(t) &=& e^{-\frac{t^2}{2\sigma_{\text{pu}}^2}}\\
a_{\text{pr}}(t) &=& e^{-\frac{(t-\tau)^2}{2\sigma_{\text{pr}}^2}}
\end{eqnarray}
\noindent $\tau$ is the delay between the X-ray probe pulse
and the optical pump that initiates a chiral dynamics, see Fig. \ref{fig1}(b).
The signal measured by spectrally dispersing the probe depends on the dispersed frequency $\omega$ and the pump-probe time delay $\tau$. The time and frequency resolved absorption of a weak probe $A_{\text{L/R}}$ is given by\cite{alexandre2002third,mukamel1} :
\begin{eqnarray}
A^{\text{L/R}}(\omega, \tau) &=& 2 \omega \Im\Big( \bold E^{\text{L/R}*}_{\text{pr}}(\omega)\cdot\bold P^{\text{L/R}}(\omega,\tau) + \bold B^{\text{L/R}*}_{\text{pr}}(\omega)\cdot\bold M^{\text{L/R}}(\omega,\tau)\Big)
\label{sigdef1}
\end{eqnarray}
\noindent where $\bold P^{\text{L/R}}(\omega,\tau)$ and $\bold M^{\text{L/R}}(\omega,\tau)$ are the $\omega$ Fourier component of the polarization ($\bold P(t) = \langle \bm \mu \rangle$) and magnetization ($\bold M(t) = \langle \bold m \rangle$) respectively.
The time and frequency resolved TRCD signal is given by :
\begin{eqnarray}
S_{\text{TRCD}}(\omega,\tau) = 2 \omega \Im \Big( &&\bold E^{\text{L}*}_{\text{pr}}(\omega)\cdot\bold P^{\text{L}}(\omega,\tau) + \bold B^{\text{L}*}_{\text{pr}}(\omega)\cdot\bold M^{\text{L}}(\omega,\tau)\nonumber\\
-&&\bold E^{\text{R}*}_{\text{pr}}(\omega)\cdot\bold P^{\text{R}}(\omega,\tau)- \bold B^{\text{R}*}_{\text{pr}}(\omega)\cdot\bold M^{\text{R}}(\omega,\tau)\Big)
\label{signaldef}
\end{eqnarray}

\begin{figure}[!h]
  \centering
  \includegraphics[width=1.\textwidth]{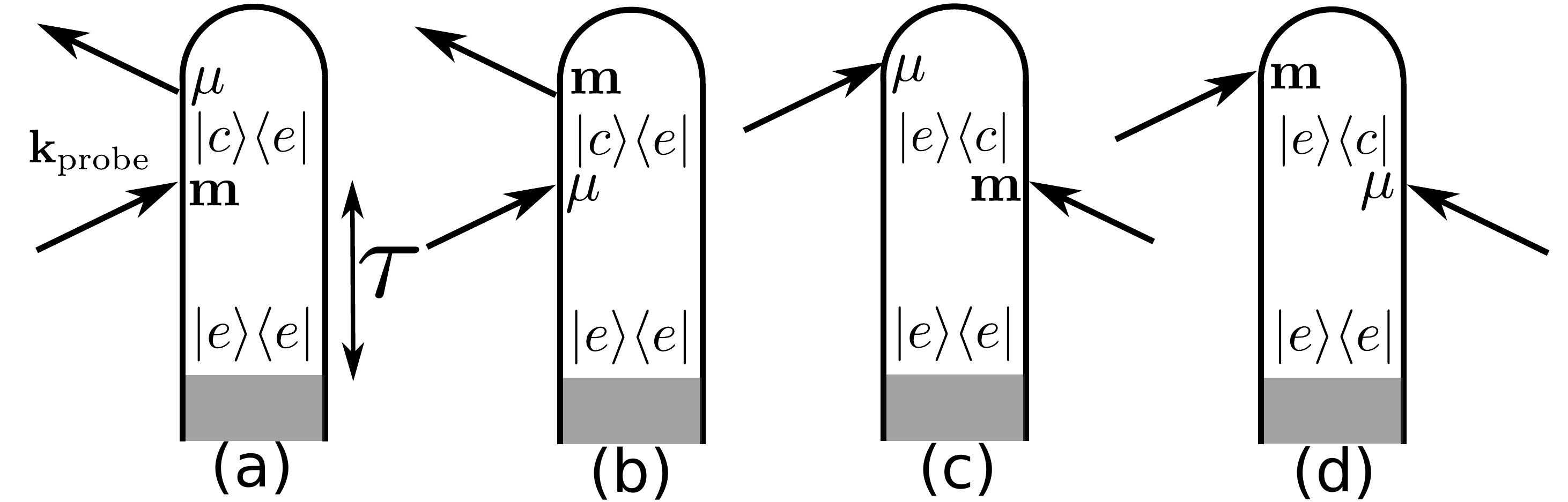}
  \caption{Loop diagrams\cite{Mukamel2010223} contributing to the TRCD signal defined in equation \ref{signaldef}. One has to consider the interaction with both the $\bm \mu \cdot \bold E$ or $\bm m \cdot \bold B$ parts of the Hamiltonian. The complex conjugates of these diagrams also contribute to the signal. Diagrams (a) and (b) represent stimulated Raman, (c) and (d) represent excited state absorption. Arrows represent the interactions with the probe. The interaction with the pump is treated implicitly and occurs during the shaded area.
  \label{fig1b}}
\end{figure}

Only the pseudo-scalar quantity of the signal, that contains one interaction with the electric dipole and one with the magnetic dipole, survives rotational averaging in Eq. \ref{signaldef} and the signal vanishes in the dipole approximation. 
The interaction with the X-ray probe is calculated perturbatively in $H_{\text{pr}}(t)$. The interaction with the pump is treated non-pertubatively and included directly in the propagator $U$ of the system as described in the next section.
Expanding the polarization in Eq. \ref{signaldef} to first order in the probe field leads to :
\begin{eqnarray}
S_\text{TRCD}(\omega,\tau) = && \frac{2}{\hbar}\omega\Re  \int_{-\infty}^{+\infty} dt e^{-i \omega t}\int_{0}^t dt_1 \nonumber\\
&& \langle \Psi_0 | U^\dagger(t,0)\bm\mu^\dagger_{ec} U(t,t_1)\bold m_{ce} U(t_1,0)|\Psi_0\rangle_{\Omega} \ \bold E^{L*}_{\text{pr}}(\omega)\cdot\bold B^L_{\text{pr}}(t_1,\tau)\nonumber\\
&+& \langle \Psi_0 | U^\dagger(t,0)\bm m_{ec}^\dagger U(t,t_1)\bm \mu_{ce} U(t_1,0)|\Psi_0\rangle_{\Omega} \ \bold B^{L*}_{\text{pr}}(\omega)\cdot\bold E^L_{\text{pr}}(t_1,\tau)\nonumber\\
&-& \langle \Psi_0 | U^\dagger(t_1,0)\bm m_{ce}^\dagger U^\dagger(t,t_1)\bm \mu_{ce} U(t,0)|\Psi_0\rangle_{\Omega}  \ \bold E^{L*}_{\text{pr}}(\omega)\cdot\bold B^{L*}_{\text{pr}}(t_1,\tau)\nonumber\\
&-& \langle \Psi_0 | U^\dagger(t_1,0)\bm\mu_{ce}^\dagger U(t,t_1)\bold m_{ce} U(t,0)|\Psi_0\rangle_{\Omega} \ \bold B^L_{\text{pr}}(\omega)\cdot\bold E^{L*}_{\text{pr}}(t_1,\tau)\nonumber\\
&-& L \longleftrightarrow R
\label{signaldef2}
\end{eqnarray}
\noindent The four terms correspond respectively to the 4 loop diagrams in Fig. \ref{fig1b}. $U(t_2,t_1)$ is the time evolution operator between times $t_1$ and $t_2$ governed by $H_0 + H_\text{pu}$ and $|\Psi_0\rangle$ is the matter ground state wavefunction. $L \longleftrightarrow R$ represents the same terms as the first 5 lines of Eq. \ref{signaldef2} with a right polarization instead of a left one.
$\langle ...\rangle_\Omega$ stands for rotational averaging over the material quantities. Rotational averaging of second rank cartesian tensor leads to
\begin{equation}
\langle\bold T\rangle_\Omega = \frac{1}{3} \mathbb{1} \text{Tr}\bold T 
\end{equation}
\noindent where $\mathbb{1}$ is the identity matrix.
Equation \ref{signaldef2} can be simplified using the standard definition for the circular polarization vectors ($\bold e_L = 1/\sqrt 2 \ (-1,i,0)$, $\bold e_R = 1/\sqrt 2 \ (1,i,0)$, $\bold b_L = 1/\sqrt 2 \ (-i,-1,0)$ and  $\bold b_R = 1/\sqrt 2 \ (-i,1,0)$\cite{varshalovich}).
We define the electric-magnetic, the magnetic-electric, and the electric-electric response functions by
\begin{eqnarray}
\bold R_{em}(t,t_1) &=& \langle \Psi_0 | U^\dagger(t,0)\bm\mu^\dagger_{ec} U(t,t_1)\bold m_{ce} U(t_1,0)|\Psi_0\rangle_{\Omega}\\
\bold R_{me}(t,t_1) &=& \langle \Psi_0 | U^\dagger(t,0)\bm m^\dagger_{ec} U(t,t_1)\bm \mu_{ce} U(t_1,0)|\Psi_0\rangle_{\Omega}\\
\bold R_{ee}(t,t_1) &=& \langle \Psi_0 | U^\dagger(t,0)\bm \mu^\dagger_{ec} U(t,t_1)\bm \mu_{ce} U(t_1,0)|\Psi_0\rangle_{\Omega}
\end{eqnarray}
The  time and frequency resolved X-ray circular
dichroism signal Eq. \ref{signaldef2} is finally given by
\begin{equation}
S_\text{TRCD}(\omega,\tau) = -\frac{4}{\hbar}\Im\int dt_1 \ \frac{1}{3}\text{Tr} [\bold R_{em}(\omega,t_1)-\bold R_{me}(\omega,t_1)] A(\omega)A(t_1-\tau)
\label{finalSwt}
\end{equation}
$\bold R_{ee}(t,t_1)$ does not contribute to the rotationally averaged signal.
We shall also consider the time-resolved (frequency-integrated) signal
\begin{equation}
S_\text{TRCD}(\tau) = \int d\omega S_\text{TRCD}(\omega,\tau)
\label{finalSt}
\end{equation}
As a reference we also present the ordinary non-chiral pump-probe signal calculated by considering only the electric-electric contribution.
\begin{equation}
S_\text{PP}(\omega,\tau) = -\frac{4}{\hbar}\Re\int dt_1 \ \frac{1}{3}\text{Tr}\bold R_{ee}(\omega,t_1)A(\omega)A(t_1-\tau)
\label{finalPP}
\end{equation}

\section{Application to the C, N and O K-edges of formamide}
\label{section3}

\begin{figure}[!h]
  \centering
  \includegraphics[width=0.7\textwidth]{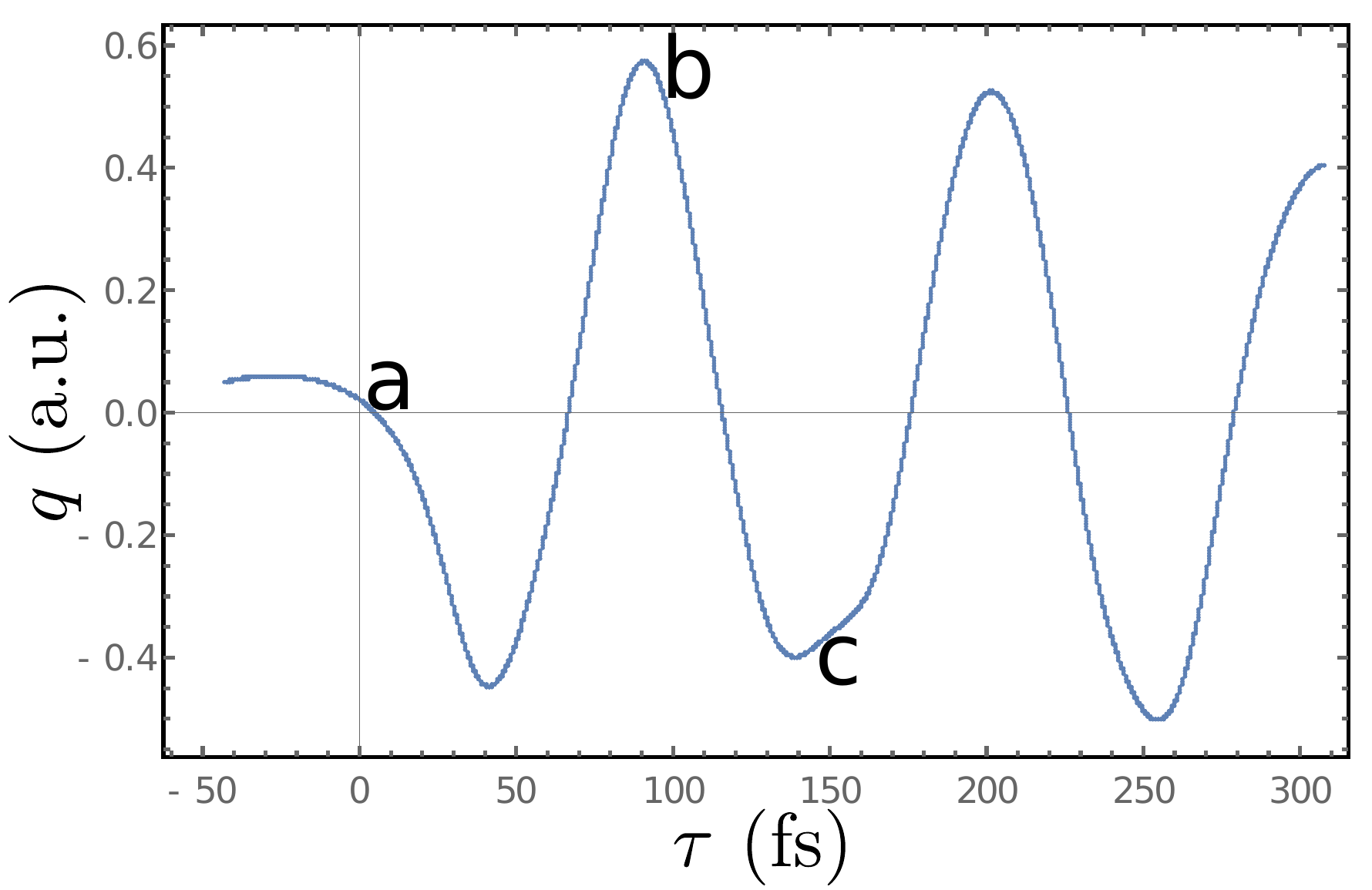}
  \caption{Bending dynamics following a 30 fs, 5.55 eV pump excitation. Shown is the time dependent average $q$.
The corresponding molecular geometries a, b and c of Fig. \ref{fig0} are marked at three key points.\label{fig3}}
\end{figure}
A left polarized pump-pulse $\bold e_L$ creates an enantiomer excess in the excited state,
by localizing the wave packet to the left side of the double well potential (negative $q$).
We assume a planar configuration in the ground state to account for the fact that
the molecule is achiral due to the low inversion barrier of the NH$_2$ group.
We select the normal mode at 1170\,cm$^{-1}$, which corresponds to the out-of-plane bending of the CHO group. The planar geometry is then displaced by the eigenvector of the selected normal mode displayed in Fig. \ref{fig1}(c) associated with thee bending motion. For each displacement step (steps of 0.05 of the displacement unit vector), the valence and core excited state are calculated using CASSCF as described in the App. \ref{app:QC}, leading to the potential energy surfaces $V_i(q)$ of the valence and core excited states presented in Fig. \ref{fig1}(a), where $i$ is either the ground state $g$, the valence
excited state $e$ or one of the core excited states $c$.

Before the pump arrival, the molecule initial state $\ke{\Psi_0} = \ke{\phi_0} \otimes \ke{g}$ is set as the vibrational ground state $\ke{\phi_0}$ in the electronic ground state $\ke{g}$.  The field-free molecular Hamiltonian including the normal mode $q$ and the electronic degrees of freedom is given by
\begin{equation}
H_0 = -\frac{1}{2m}\frac{d^2}{dq^2} + V_g(q) + V_e(q) + V_c(q)
\end{equation} 
where  $m$ is the reduced mass of the mass scaled normal mode motion (1\,amu).
The time-dependent Schr{\"o}dinger equation with the Hamiltonian $H_0 + H_{\text{pu}}(t)$, Eq. \ref{hamiltonian2}, is solved numerically on a one-dimensional numerical grid (see appendix \ref{app:QD} for detailed information). 
The electric and magnetic fields used in $H_{\text{pu}}(t)$ in Eq. \ref{hamiltonian2} are left circularly polarized and have a Gaussian envelope tuned at a frequency slightly below the ground to first valence excited state transition (5.85 eV) in order to maximize the enantiomeric excess ($\sigma_{\text{pu}} = 30$ fs, $\omega_{\text{pu}} = 5.55$ eV). The rotating wave approximation has been used to remove the rapid oscillation of the carrier frequency in the propagation.
  
\begin{figure}[!h]
  \centering
  \includegraphics[width=0.7\textwidth]{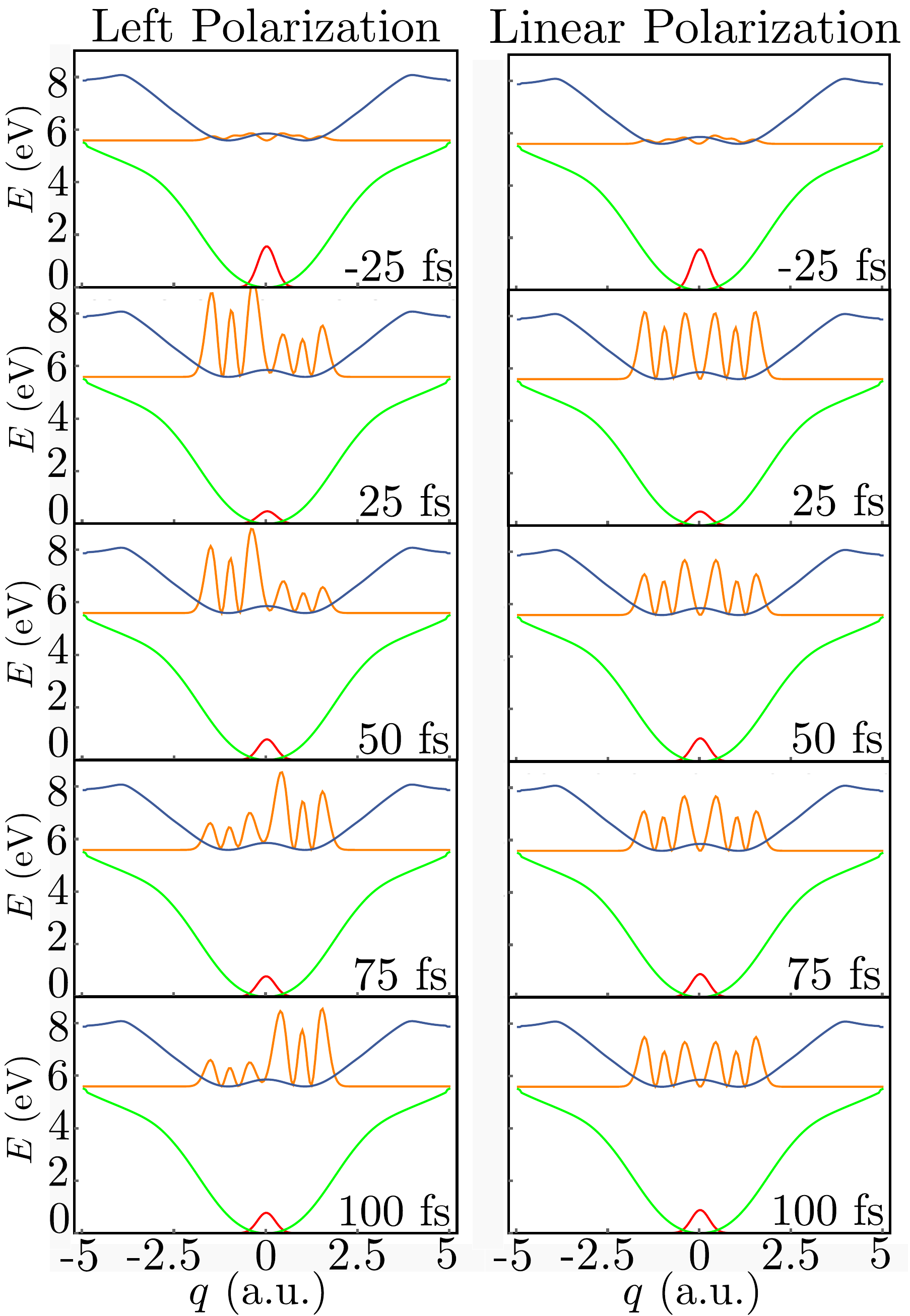}
  \caption{Left column : wavepacket dynamics induced by a left circularly polarized $\sigma_\text{pu}$30 fs, 5.55 eV pump pulse arriving at 0 fs for different delays $\tau$. Right column, same but for a linearly polarized pump that does not generate chirality. The time after the pump arrival is indicated in each panel.
  \label{fig4}}
\end{figure}

The excited state nuclear population dynamics along the out-of-plane nuclear coordinates $q$, Fig. \ref{fig1}(c), then evolves during the delay $\tau$ as shown in Fig. \ref{fig3} and the evolving nuclear wavepacket is displayed in Fig. \ref{fig4}.
As a reference, we also show the population dynamics for a linearly polarized excitation which does not create an enantiomer excess and thus does not generate a chiral signal. Animations of the dynamics for various left polarized pump duration ($\sigma_\text{pu} = $40, 20, 10, 5, 1 fs) and a linearly polarized pump  ($\sigma_\text{pu} = $30 fs) is given in supplementary materials.

At each molecular geometry along the dynamics, the first valence and core state are calculated as described in the methods section.
The resulting lowest lying core-hole transition is 286 eV (282 eV experimentally\cite{edgesVALUES}) for the C K-edge, 400 eV for the N K-edge (397 eV experimentally\cite{edgesVALUES}) and 529 eV for the O K-edge (533 eV experimentally\cite{edgesVALUES}).

\begin{figure}[!h]
  \centering
  \includegraphics[width=1.1\textwidth]{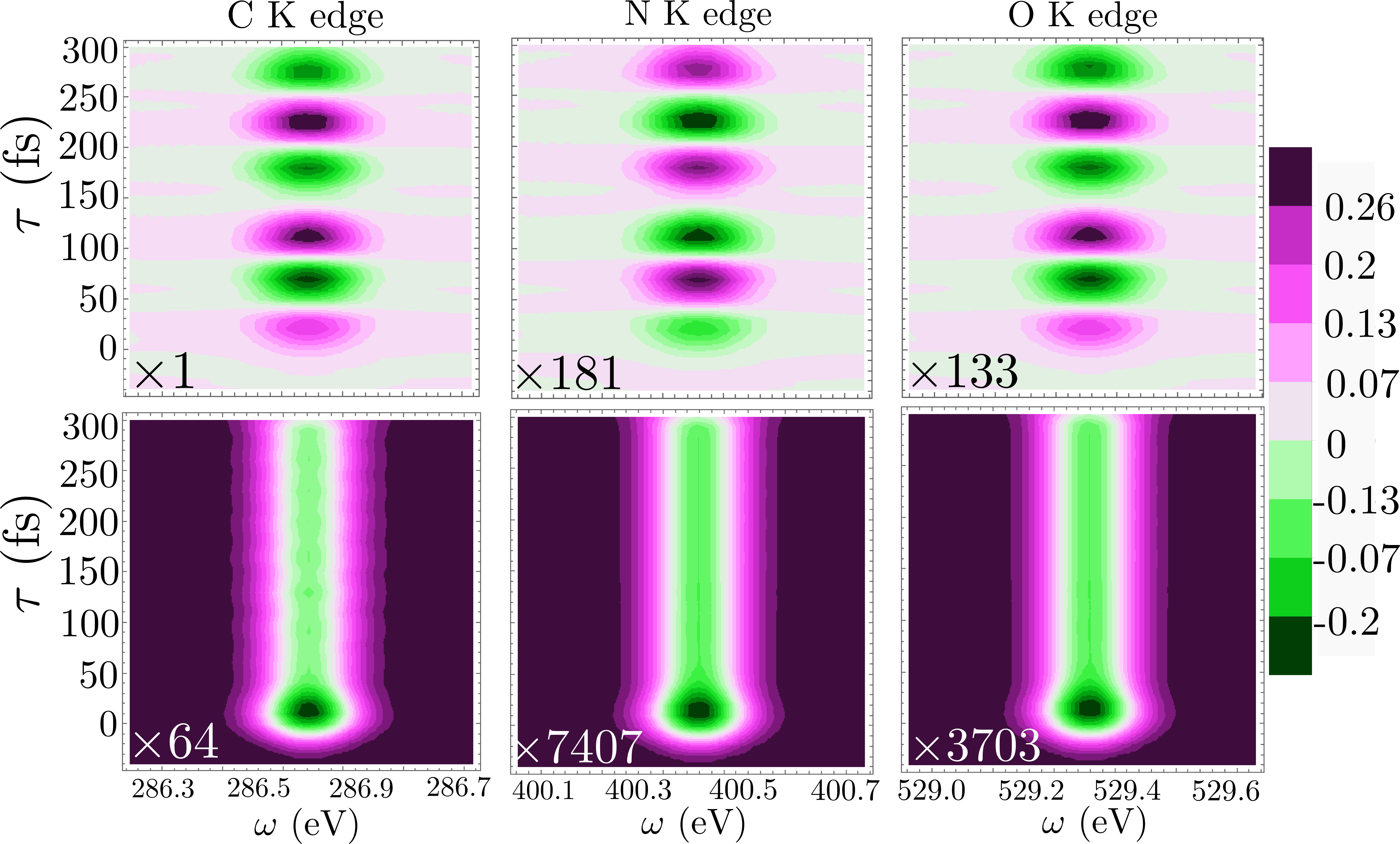}
  \caption{Top row : frequency and time resolved CD, Eq. \ref{finalSwt}, at the C, N and O K edges of formamide. Bottom row : corresponding frequency resolved pump-probe non-chiral signals, Eq. \ref{finalPP} at the C, N and O K edges. The color bar indicates the amplitude of the C K-edge CD signal. The spectra are normalized on the same scale and their absolute magnitude is multiplied by the factor on the bottom left of each graph.
  \label{fig5}}
\end{figure}

The time and frequency dispersed TRCD signals for the C, N and O K-edges are displayed in Fig. \ref{fig5}, top row. The probes are respectively tuned at the valence to K-edge transition with $\sigma_{\text{pr}} = 20$ fs. The signals show an oscillatory pattern with the same period (120 fs) as the enantiomeric excess dynamics shown in Fig. \ref{fig3}. Indeed, the molecule is back to its original position at the end of a period and the X-ray light is probing the same geometry.
\begin{figure}[!h]
  \centering
  \includegraphics[width=0.5\textwidth]{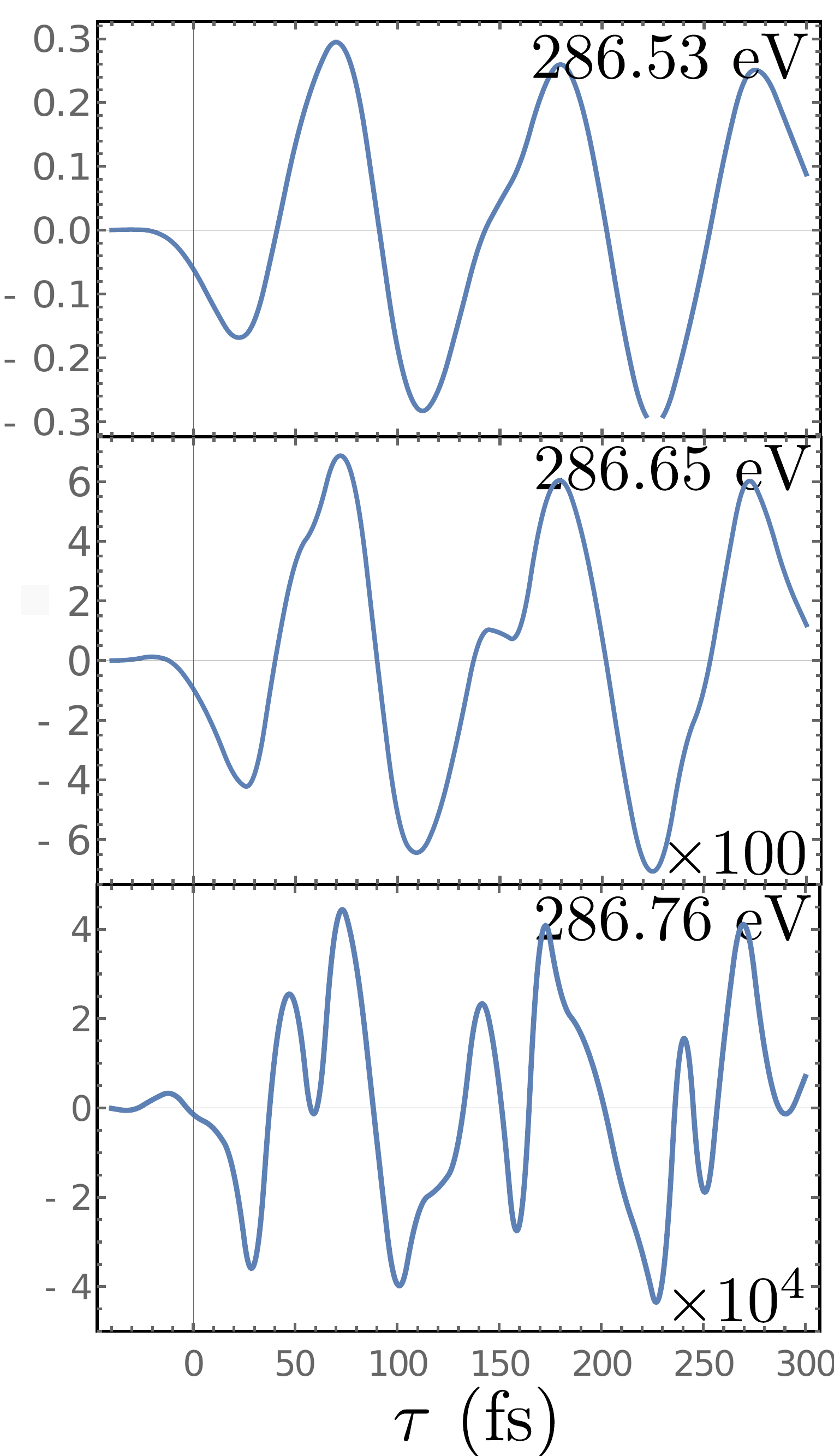}
  \caption{Vertical slices of the frequency and time resolved C K-edges chiral signal $S_\text{TRCD}(\omega,\tau)$, Eq. \ref{finalSwt}. The $\omega$ values are indicated in each panel.
  \label{fig6}}
\end{figure}
In Fig. \ref{fig5}, bottom row, we display the non-chiral pump-probe signal for a left polarized probe. This signal is insensitive to the enantiomeric excess dynamics and does not show the oscillation. It is about 2 orders of magnitude stronger than the CD signal. This is a typical relative magnitude of CD versus non chiral signals\cite{berovaCD}. From Fig. \ref{fig5}, the relative magnitudes of the CD signal compared to the non-chiral contributions are 1.6\%, 2.4\% and 3.6\%  for the C, N and O edges respectively. Vertical slices of the time and frequency resolved signal of Fig. \ref{fig5} are displayed in Fig. \ref{fig6}. 

\begin{figure}[!h]
  \centering
  \includegraphics[width=0.7\textwidth]{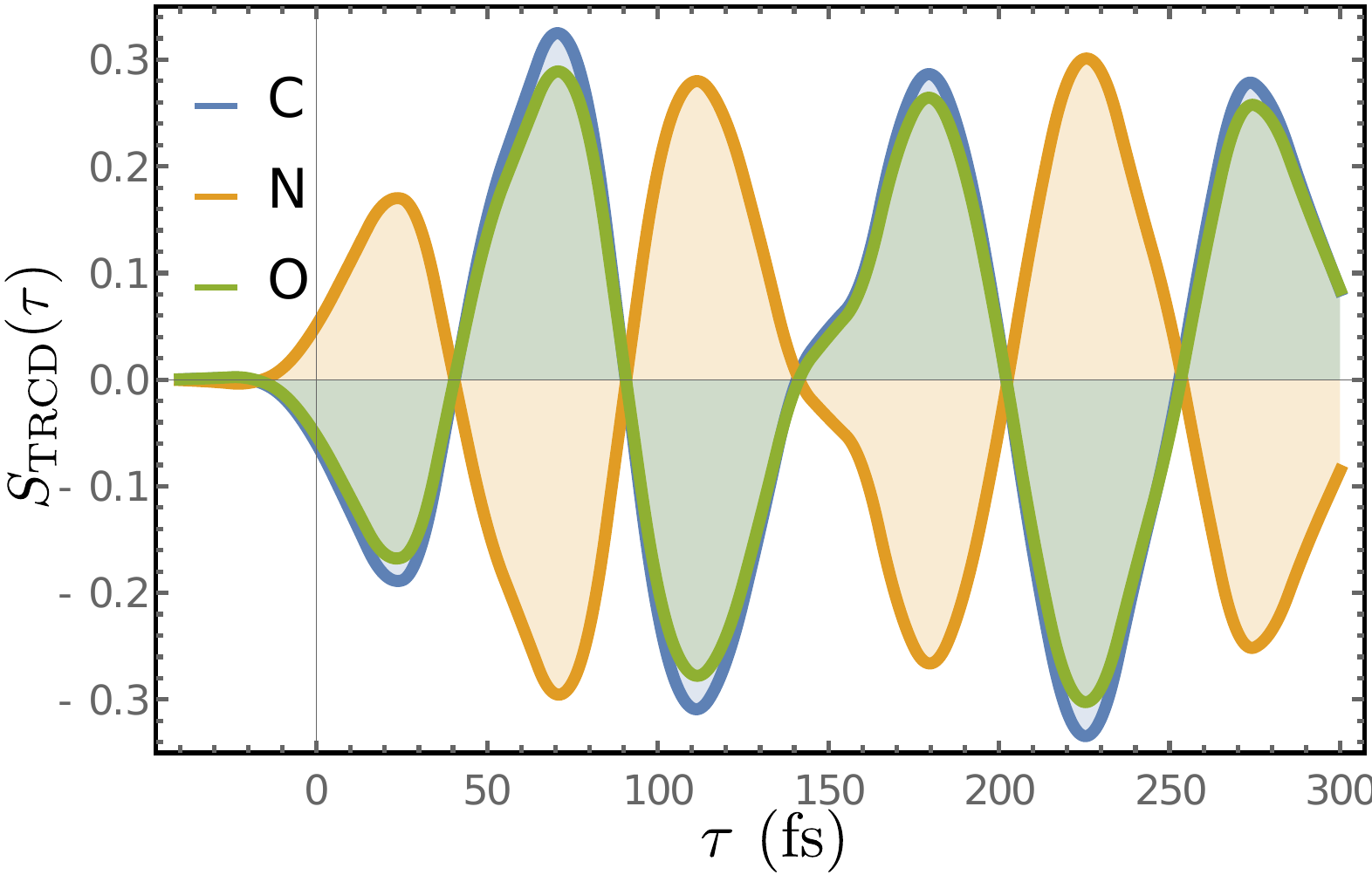}
  \caption{Time-resolved CD of formamide, Eq. \ref{finalSwt} at the C (blue), N (orange) and O (green) K edges.
  \label{fig7}}
\end{figure}

The signals for the various K-edges are very similar. This is due to the fact that the molecule is small and all cores are in close proximity to the chiral center, the C atom.
Thus, the different atoms reveal the same dynamics along the out-of-plane normal coordinate. 
The corresponding time-resolved signal, Eq. \ref{finalSt}, shown in Fig. \ref{fig7} reveals
the $\simeq 120$ fs oscillatory period. The TRCD signals closely resemble the dynamics of the expectation of the nuclear coordinate, revealing the enantiomer excess.

\begin{figure}[!h]
  \centering
  \includegraphics[width=0.5\textwidth]{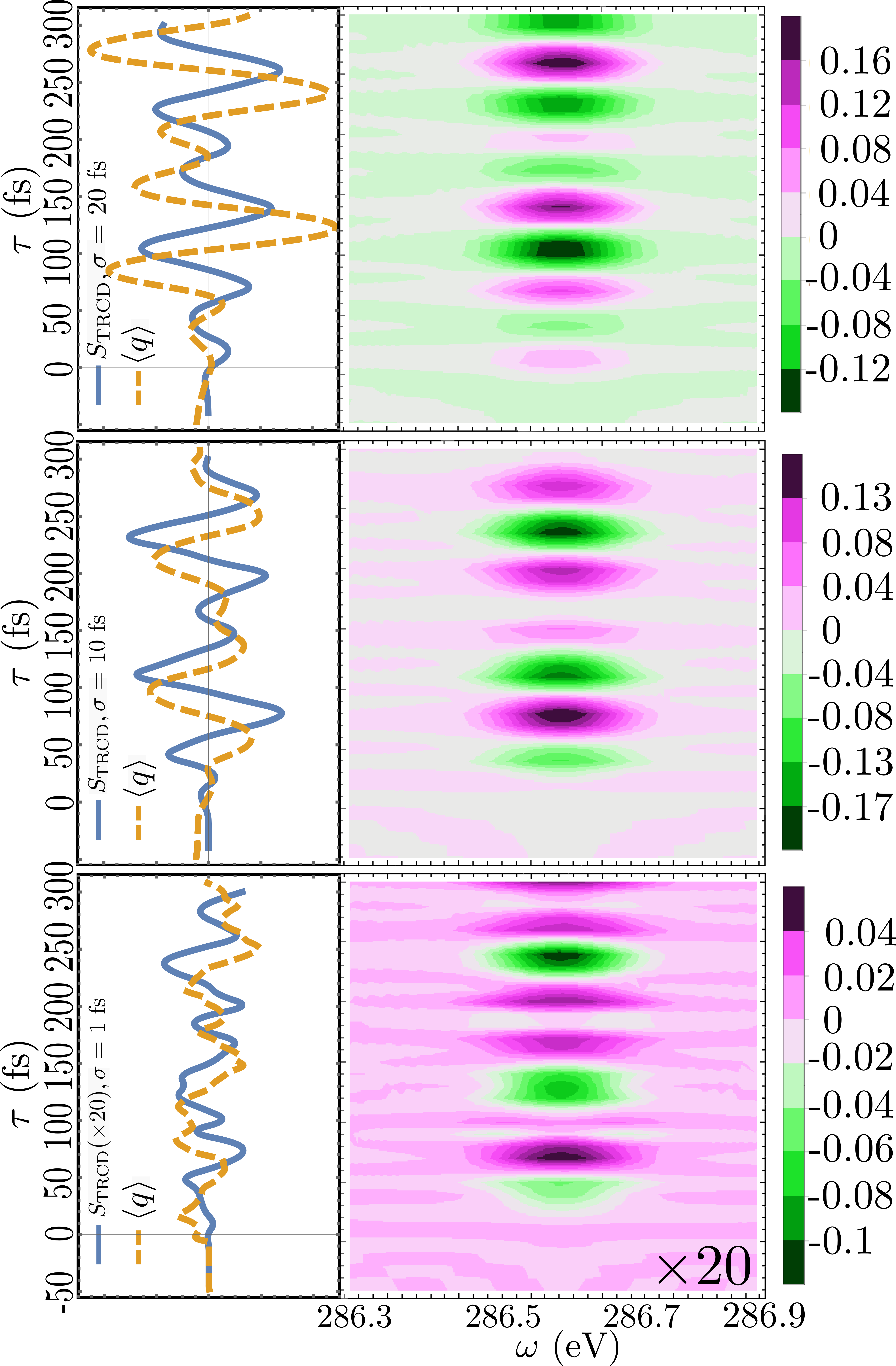}
  \caption{Left : TRCD signals at the C K-edge (solid, blue, Eq. \ref{finalSt}) and expectation value of the normal mode coordinate $q$ (dashed, orange) calculated for three pump durations : $\sigma_{\text{pu}} = $ 20, 10 and 1 fs ($\omega_{\text{pu}}$ = 5.85 eV for all). Right : Corresponding frequency and time resolved signals.
  \label{fig8}}
\end{figure}

Finally, we present in Fig. \ref{fig8} the TRCD signals at the C K-edge calculated for three pump-pulse lengths : $\sigma_{\text{pu}} = $ 20, 10 and 1 fs ($\omega_{\text{pu}}$ = 5.85 eV for all). As can be seen in the video given in supplementary material, the wavepacket dynamics depends on the pump duration as shown by the expectation value of the out-of-plane motion along $q$. The CD signals become weaker as the maximum modulation in $\langle q \rangle$ becomes smaller.

\section{Conclusions}
\label{section4}

We have demonstrated how ultrafast molecular chiral dynamics may be probed using circularly polarized X-ray pulses. 
Measuring these signals requires a optical pump, X-ray probe setup with ultrashort circularly polarized laser light.
Molecular chirality and the corresponding signals are sensitive to the conformation.
Such signals are simpler to interpret than chiral HHG signals.

The ultrafast enantiomer conversion in formamide can be monitored in real time by measuring the time-resolved CD at various K-edges.
We found no substantial differences  between the different K-edges. 
Each K-edge is associated with a single selected atom and then correspond to local positions in the molecule. For larger molecules, one can expect multiple identical atoms to contribute to the same core resonant signal and to yield more global geometric information
 In particular, one can expect to be able to probe at different structural dynamics by probing inequivalent C-K edges in larger molecules. 
For a simple molecule like formamide the excited state dynamics is dominated by a single vibrational mode but we can expect the signals to be different for larger molecule experience multiple dynamics on various timescale e.g., in protein folding.

\begin{appendices}
 %%%%%%%%%%%%%%%%%%%%%%%%%%%%%%%%%%%%%%%%%%%%%%%%%%%%%%%
 %%%%%%%%%%%%%%%%%%%%%%%%%%%%%%%%%%%%%%%%%%%%%%%%%%%%%%%
  %%%%%%%%%%%%%%%%%%%%%%%%%%%%%%%%%%%%%%%%%%%%%%%%%%%%%%% 
\section{Electronic structure simulations}
\label{app:QC}
The ground state $\ke{g}$ of formamide has been optimized on the sa3-CAS(8/8)/6-31G* level
of theory (MOLPRO \cite{MOLPRO_brief}) constrained to C$_\text{s}$ symmetry and has an imaginary normal of 100\,cm$^{-1}$. The planar geometry is used to simulate
the ground state of the double well potential along the NH$_2$ bending motion,
without having to take into account the bending motion explicitly.
The normal mode for the out-of-plane bending motion is the normal mode at 1170\,cm$^{-1}$ of the C$_\text{s}$ geometry. The valence excited $\ke{e}$ is
calculated at the same level of theory.

The core excited states are then calculated in separate RASSCF calculations,
by freezing the optimization of the 1s core orbitals of C, N and O,
rotating them into the active space and restricting their occupation to a single electron.
The different active spaces, which have been used for the calculation of the different K-edges are given in Tab. \ref{tab_cas}
\begin{table}
\caption{Active spaces for the core excited states. The frozen singly occupied core orbital is not counted in the defintion of the active space, given as (active electrons/active orbitals).}\label{tab_cas}
\begin{tabular}{lll}
\hline\hline
core orbital & active space& state average \\\hline
C(1s) & (9/6)& 2 \\
N(1s) & (7/5)& 1 \\
O(1s) & (9/7) & 1 \\
\hline\hline
\end{tabular}
\end{table}

\section{Quantum Dynamics}
\label{app:QD}
The time-evolution of the molecule including the pump-pulse is treated numerically by solving the time dependent Schr\"odinger equation on a grid and time stepping
with the Chebyshev propagation scheme \cite{Ezer84}. The interaction with
the pump-pulse is explicitly included in the propagation scheme, while the interaction
with the probe pulses is treated with perturbation theory through the calculation of the
two-time correlation functions. The respective correlation functions in Eq. \ref{finalSwt}
are then obtained by numerically propagating $\ke{\Psi_0}$ forward to $t_1$ interacting with $m_{ce}$/$\mu_{ce}$, propagating forward to $t$, interacting with $\mu^\dagger_{ce}$/$m^\dagger_{ce}$, and propagating backward to $t=0$.
\end{appendices}

\begin{acknowledgements}
The support of the Chemical Sciences, Geosciences, and Biosciences division, Office of Basic Energy Sciences, Office of Science, U.S. Department of Energy through Award No. DE-FG02-04ER15571 and of the National Science Foundation (Grant No CHE-1361516) is gratefully acknowledged. J.R. was supported by the DOE grant. M.K. gratefully acknowledges support from the Alexander von Humboldt foundation through the Feodor Lynen program.
\end{acknowledgements}

\bibliography{xrTRCD}

\end{document}